\documentclass[aps,pra,nofootinbib,twocolumn]{revtex4-2}
\usepackage[colorlinks]{hyperref}
\usepackage[utf8]{inputenc}
\usepackage{graphicx}
\usepackage{multirow}
\usepackage{amssymb}
\usepackage{amsmath}
\usepackage{gensymb}
\usepackage{xcolor,notes2bib}
\bibnotesetup{
note-name = ,
use-sort-key = false
}
\begin{document}

\title{Scalar dark matter induced oscillation of permanent-magnet field}

\author{I.~M.~Bloch$^{1}$}\email{itayblochm@berkeley.edu}
\author{D.~Budker$^{2}$}\email{budker@uni-mainz.de}
\author{V.~V.~Flambaum$^{3}$}\email{v.flambaum@unsw.edu.au}
\author{I.~B.~Samsonov$^3$}\email{igor.samsonov@unsw.edu.au}
\author{A.~O.~Sushkov$^4$}\email{asu@bu.edu}
\author{O.~Tretiak$^2$}\email{oleg.tretiak@uni-mainz.de}
\affiliation{$^1$Berkeley Center for Theoretical Physics, University of California, Berkeley, CA 94720, U.S.A. and\\
Theoretical Physics Group, Lawrence Berkeley National Laboratory, Berkeley, CA 94720, U.S.A.}
\affiliation{$^2$Johannes Gutenberg-Universit\"at Mainz, 55128 Mainz, Germany and\\
Helmholtz-Institut, GSI Helmholtzzentrum f\"ur Schwerionenforschung, 55128 Mainz, Germany}
\affiliation{$^3$School of Physics, University of New South Wales, Sydney 2052, Australia}
\affiliation{$^4$Department of Physics, Boston University, Boston, MA 02215, USA}

\begin{abstract}
Scalar-field dark matter models imply small oscillations of fundamental constants. These oscillations could result in observable variations of the magnetic field in a permanent magnet. We propose an experiment for detection of this type of dark matter through searches of oscillations of magnetic field of permanent magnets with a SQUID magnetometer or a low-noise radiofrequency amplifier. We show that this experiment may have comparable sensitivity to leading experiments searching for variations of fundamental constants in the range of frequencies from a few Hz to about 1\,MHz. We also discuss applicability of the approach of variations of fundamental constants for accounting for the interaction with scalar dark matter.
\end{abstract}

\maketitle

\section{Introduction}
Despite several decades of concerted experimental efforts, nongravitational interactions of dark matter (DM) are yet to be unambiguously detected, leaving identifying the nature of DM as one of the greatest challenges in modern science \cite{bertone2010}. Ultralight bosonic dark matter (UBDM) has emerged as a promising class of candidates \cite{kimball2022search}.

In contrast to heavier particles that may constitute DM, UBDM is searched for through its collective effects, rather than using particle detectors. Detection approaches may vary depending on the spin and intrinsic parity of the underlying particles. In the case of scalars, the potentially observable signatures may be produced by apparent modification of fundamental constants \cite{oscillation,Arvanitaki2015,Antypas2020}. 

The most notable effects are related to variations of the fine structure constant and masses of elementary particles, because they may be, in principle, observed in a variety of experiments, see, for example, \cite{Snowmass}. Since the variation of mass and charge of a single particle is extremely small, it is advantageous to look for collective effects, when an ensemble of polarized particles interacts coherently with the classical oscillating scalar field. In this case, the observable effects are enhanced through the large number of particles involved. 

In this paper, we study the effect of oscillation of magnetic field of a permanent magnet due to the interaction with a background scalar field associated with the local dark matter density. An experimental realization could measure small variations of the magnetic field with a sensitive magnetometer such as a superconducting quantum interferometer device (SQUID)~\bibnote{Such an experiment is currently in progress at Boston University.}. To provide interpretation of the experiment, one needs to find the relation between the  magnetic field in the magnet and variations of the fundamental constants. This question is the focus of this paper.

We derive the dependence of the oscillating magnetic field in a magnetized material on the coupling constants of the scalar field to photons and electrons. This field is found as the analytical solution of Maxwell's equations in a long cylindrical magnet. We show that magnetic materials with low electric conductivity are more suitable for the detection of variations of fundamental constants because eddy currents suppress the oscillating fields. The oscillating  electric and magnetic
fields may be detected with a SQUID magnetometer or with an induction coil connected to a low-noise radiofrequency (rf) amplifier. We estimate the sensitivity of both these detectors and compare these with the results of other experiments which are sensitive to variations of the fundamental constants \cite{HQuartzSapphire,DynamicDecoupling,Holometer,I2,Damned,CsCav,DyQuartz,GEO600,FifthForce,EotWash,Microscope}. We show that the proposed experiment may have comparable sensitivity to the leading experiments searching for variations of fundamental constants in the range of frequencies from a few Hz to about 1\,MHz. 

The rest of the paper is organized as follows. In the next section, we study the dependence of the magnetization of a magnetic material on the coupling constants of the scalar field to photons and electrons. We show that the electron coupling constant may be conveniently taken into account through the variation of the Bohr magneton, while the scalar-photon interaction leads to an extra term in Maxwell's equations. In Sec.\,\ref{SecMagnet}, we analytically derive the oscillating electric and magnetic fields in an axially-magnetized long cylinder, originating from the interaction with the background scalar field. Using this analytic solution, we study the effects of suppression of these fields in permanent magnets with finite electric conductivity and resonance enhancements in non-conducting magnets. In section \ref{SecSensitivity}, we estimate the sensitivity of experimental setups based on the SQUID magnetometer and low-noise rf amplifier, and compare these with the existing limits from other experiments. Sec.\,\ref{SecSummary} is devoted to a summary and a discussion of the results.


\section{Oscillating magnetization and variation of fundamental constants}
In this section, we start by considering the classical oscillating scalar field as the dark matter candidate and, then, derive variations of the Bohr magneton and magnetization due to the electron interaction with this field.

\subsection{Classical scalar field as a dark matter candidate}

We consider a model of dark matter described by a real scalar field $\phi$ with the mass of the underlying particle $m_\phi$. If this mass is sufficiently low, $m_\phi\ll 1$\,eV, there are many dark matter particles per de Broglie wavelength, and this field may be well approximated by a classical field described by a plane wave with angular frequency $\omega\approx m_\phi$,
\begin{equation}
    \phi = \phi_0 \cos(\omega t +\varphi)\,,
    \label{phi}
\end{equation}
where $\varphi$ is a location-dependent phase. Assuming that this scalar field saturates all local dark matter density $\rho_{\rm DM}= 0.4\,\text{GeV/}\text{cm}^3$, the amplitude of the plane wave (\ref{phi}) is (here we use natural units with $\hbar=c=1$)
\begin{equation}
    \phi_0 = \sqrt{2\rho_{\rm DM}}/m_\phi\,.
    \label{phi0}
\end{equation} 

In this paper, we assume that the de Broglie wavelength of the scalar field is much larger than the experimental setups used for its detection. Furthermore, we focus on experiments with the integration time much longer than $1/\omega$, which are sensitive to many oscillations. Under these assumptions, the phase in Eq.\,(\ref{phi}) does not play an important  role, and, for simplicity, we set $\varphi=0$.

Our calculations are performed for the plane wave  (\ref{phi})  with definite frequency $\omega$ since the spread of frequencies is relatively small. For example, in the standard halo model of dark matter   $\delta \omega \simeq 10^{-6}\omega$. 

\subsection{Can we replace  the interaction with the scalar field by oscillating fundamental constants?}

The interaction of the scalar field (\ref{phi}) with the electromagnetic field $F_{\mu\nu}$ and a Dirac electron $\Psi$ with mass $m_e$ is described by the Lagrangian
\begin{equation}
    {\cal L}_{\rm int} = \frac14 g_\gamma \phi
    F_{\mu\nu}F^{\mu\nu} - g_e \phi \bar\Psi \Psi\,,
    \label{L}
\end{equation}
where $g_\gamma$ and $g_e$ are coupling constants. This Lagrangian is similar to the free Lagrangian for the electromagnetic field and the fermion mass term,
\begin{equation}
    {\cal L} = - \frac14
    F_{\mu\nu}F^{\mu\nu} - m_e \bar\Psi \Psi\,.
    \label{Lfree}
\end{equation}
Therefore, it is convenient to take into account the fermion interaction term in Eq.\,(\ref{L}) by a redefinition of the electron mass, $m_e\to m_e' = m_e + \delta m_e$ with
\begin{equation}
    \delta m_e = g_e \phi_0 \cos(\omega t)\,.
    \label{deltame}
\end{equation}
This oscillating effective electron mass $m_e'$ should imply observable effects such as oscillating magnetization which we consider in this paper.

The first term in the right-hand side in Eq.\,(\ref{L}) yields a modification of the photon propagator when the space-time derivatives of $\phi$ may be ignored. Alternatively, it can be taken into account by a redefinition of the fine structure constant, $\alpha\to \alpha+\delta\alpha$, with $\delta\alpha/\alpha = g_\gamma \phi$ \cite{oscillation}. Indeed, the electromagnetic interaction between fermions contains a product of $e^2$ and the photon propagator $1/[q^2 (1 - g_\gamma \phi)]$, where $q$ is the momentum transfer. Therefore, this modification of the photon propagator may be fully accommodated by the corresponding change of the electron charge $e$ and the fine structure constant $\alpha=e^2$. However, this must be taken with care, as only pairs of charges should be varied which are connected by the photon propagator, and the photon propagator should involve no scalar field to avoid double counting of this interaction.

Consider, for example, an effect of oscillating magnetic field due to scalar dark matter. Naively, one may conclude that this effect is first-order in variation of the electron charge $e$ since the source of magnetic field, the electron magnetic moment, proportional to the Bohr magneton  $\mu_B = e/(2m_e)$, is first order in  $e$.  However, to have a consistent result using the variation-of-$\alpha$ approach, one must add the effect of the variation of $e$ in the detector, and the total effect contains a variation of $e^2=\alpha$. 

This result may be obtained in a different way.  Instead of variation of $\alpha$ we consider the variation of the magnetic field $B$ due to the interaction with the scalar field (\ref{phi}). Both the variation of $\alpha$ and $B$ originate from the first term in Eq.\,(\ref{L}), so adding these two contributions would be double counting. The  variation of the magnetic field $\delta B = g_\gamma B_0 \phi_0\cos(\omega t)$ is equivalent to the variation of $\alpha=e^2$ rather than the variation of $e$.

In this paper, it is  convenient to take the following approach. We consider the oscillating electron mass (\ref{deltame}), but keep the electron charge $e$ constant, while including the interaction between the  scalar field and electromagnetic field explicitly. The interaction with the coupling constant $g_\gamma$ in Eq.\,(\ref{L}) will result in an extra term in Maxwell equations (see Eqs.\,(\ref{Maxwell}) in the next section).

\subsection{Oscillating Bohr magneton and magnetization}

The electron magnetic moment should oscillate as a consequence of the oscillating electron mass (\ref{deltame}). Indeed, the definition of the Bohr magneton $\mu_B = e/(2m_e)$ implies
\begin{equation}
    \frac{\delta \mu_B}{\mu_B} = - \frac{\delta m_e}{m_e}
    = -\frac{g_e}{m_e}\phi_0 \cos(\omega t)\,.
\end{equation}
The magnetization $\vec M$ of a permanent magnet occurs mainly due to orientation of angular momenta of bound valence electrons. Thus,
\begin{equation}
    \frac{\delta M }{M_0} = \frac{\delta \mu_B}{\mu_B} = -\frac{g_e}{m_e} \phi_0\cos(\omega t)\,,
    \label{deltaM}
\end{equation}
where $M_0$ is the permanent magnetization of the magnet. 

More generally, Eq.\,(\ref{deltaM}) should involve also the variation of the magnetic $g$ factor which accounts for both spin $S$ and orbital angular momentum $L$ contributions. However, the effect of variation of $g$ is significantly smaller than the effect of variation of $\mu_B$. Indeed, interaction between atomic electron and scalar field perturbs electron wave function, $\psi=\psi_0 + \delta \psi$, with $\delta \psi$ having the same angular quantum numbers as $\psi_0$. In the non-relativistic approximation, for an isolated atom the non-diagonal matrix element of the magnetic moment operator vanishes, $\mu_B\langle \delta \psi | L_z +2 S_z|\psi_0\rangle =0$, because of the orthogonality of the radial electron wave functions with different principal quantum numbers $n$ and the same angular quantum numbers, hence, $\langle \psi | L_z +2 S_z|\psi\rangle  \approx \langle \psi_0 | L_z +2 S_z|\psi_0\rangle$. Thus, for an isolated atom  corrections to the $g$-factor due to the scalar dark matter are small and may be ignored. We assume that this conclusion holds approximately for an atom in a solid. 

Naively, the oscillating contribution to the magnetic field of the long cylindrical magnet due to the interaction with the scalar field dark matter is 
\begin{equation}
\delta B = 4\pi \delta M\,,
\label{deltaB}
\end{equation}
where $\delta M$ is given in Eq.\,(\ref{deltaM}). As we will show, Eq.\,(\ref{deltaB}) is indeed correct for magnets with low electric conductivity and when the de Broglie wavelength of the scalar field is much larger than the magnet size. In general, however, the oscillating magnetization (\ref{deltaM}) should be considered as {\it a source} in the Maxwell equations describing dynamics of electric and magnetic fields of the magnet. We stress the importance of deriving the resulting oscillating fields of the magnet as solutions of the Maxwell equations, because these fields should obey correct boundary conditions, and screening effects from eddy currents in the magnet should be considered.


\section{Oscillating electric and magnetic fields in a magnet}
\label{SecMagnet}

In this section, we derive electric and magnetic fields as a solution of the Maxwell equations sourced by the oscillating magnetization. This solution is then applied to study screening effects due to eddy currents in magnets with high electric conductivity and resonance enhancement effects in non-conducting magnets.

\subsection{Maxwell equations with oscillating magnetization}

In general, in a medium described by a relative permittivity $\epsilon$ the propagation of the electric field strength $\vec E$ and magnetic flux density $\vec B$ are described by the Maxwell equations,
\begin{subequations}
\label{Maxwell}
\begin{align}
    &\nabla\cdot \vec D =0\,,\qquad \nabla\cdot \vec B=0\,,\\
    &\nabla\times \vec E = -\frac1c \frac{\partial \vec B}{\partial t}\,,\\
    &\nabla\times (\vec H - g_\gamma \phi \vec B) = \frac{4\pi}{c}\vec j + \frac1c \frac{\partial \vec D}{\partial t}\,,
\label{Maxwell-c}
\end{align}
\end{subequations}
where $\vec D = \epsilon \vec E$ is the electric displacement field and $\vec H = \vec B - 4\pi \vec M$ is the magnetic field strength, $\vec M$ is the magnetization vector, and $\vec j$ is the current density. Equation (\ref{Maxwell-c}) includes also the term with explicit interaction of the scalar field $\phi$ originating from the first term in the Lagrangian (\ref{L}). In Eqs.\,(\ref{Maxwell}), we have neglected the term proportional to $g_\gamma \phi {\vec E}$, because in a permanent magnet this term is higher-order with respect to the coupling $g_{\gamma}$.

With no scalar field dark matter, a magnet possesses a permanent magnetization $\vec M = \vec M_0$ that produces the permanent magnetic field $\vec B_0$ (which is $\vec B_0=4\pi \vec M_0$  inside a long cylindrical magnet). The interaction of electrons with the scalar field dark matter, however, yields the oscillations of the magnetization (\ref{deltaM}). It is therefore convenient to decompose the total magnetization and the magnetic flux density as
\begin{align}
    \vec M &= \vec M_0 +  \delta\vec M\,,\\
    \vec B &= \vec B_0 + \delta \vec B\,,
\end{align}
where $\delta\vec M$ is given by Eq.\,(\ref{deltaM}), and $\delta \vec B$ obeys the following corollary of the Maxwell equations (\ref{Maxwell}):
\begin{subequations}
\label{MaxwellB1}
\begin{align}
    &\nabla\cdot (\epsilon\vec E) =0\,,\qquad \nabla\cdot \delta\vec B=0\,,\\
    &\nabla\times \vec E = -\frac1c \frac{\partial \delta\vec B}{\partial t}\,,\\
    &\nabla\times \delta \vec B =\frac{4\pi}{c}(\vec j + \vec j_{\rm eff}) + \frac1c \frac{\partial (\epsilon\vec E)}{\partial t}\,.
\end{align}
\end{subequations}
Here 
\begin{equation}
\vec j = \sigma \vec E    
\end{equation}
is the physical current density in the medium with electric conductivity $\sigma$ and
\begin{equation}
    \vec j_{\rm eff} = c g \phi \nabla\times \vec M_0
\label{jeff}
\end{equation}
is the effective current density corresponding to the oscillating magnetization (\ref{deltaM}) and 
\begin{equation}
    g = g_\gamma - \frac{g_e}{m_e}
    \label{geff}
\end{equation}
is the combination of coupling constants which is relevant for our setups.

The physical solutions of Eqs.\,(\ref{MaxwellB1}) obey the boundary conditions on the boundary between the two media:
\begin{equation}
\begin{aligned}
    &\vec D_{n[1]} = \vec D_{n[2]}\,,\quad 
     \vec E_{t[1]} = \vec E_{t[2]}\,,\\
    &\vec H_{t[1]} = \vec H_{t[2]}\,,\quad
     \vec B_{n[1]} = \vec B_{n[2]}\,,
\end{aligned}
\end{equation}
where the subscripts `t' and `n' stand for the tangential and normal components to the boundary, respectively.

Note that, in general, both relative permittivity $\epsilon$ and electric conductivity $\sigma$ contain oscillating contributions originating from the interaction with the scalar field dark matter. In Eqs.\,(\ref{MaxwellB1}), these oscillating terms in $\epsilon$ and $\sigma$ may be ignored, because they correspond to higher-order corrections with respect to the interaction constant $g$. Therefore, in what follows, we will consider $\epsilon$ and $\sigma$ independent of $g$ and constant in time.

\subsection{Infinite cylindrical magnet}

Consider an infinite cylindrical magnet of radius $R$ aligned along the $z$-axis. Assume that the magnetization vector $\vec M_0=(0,0,M_{0z})$ has the only non-vanishing component along this axis,
\begin{equation}
    M_{0z} = M_0 \theta(R-r)\,,
\end{equation}
where $\theta(R-r)$ is the Heaviside step function in cylindrical coordinates $(r,\varphi,z)$. With this expression for the magnetization, we find the effective current (\ref{jeff}) to be $\vec j_{\rm eff} = (0,j_{{\rm eff},\varphi},0)$, where
\begin{equation}
    j_{{\rm eff},\varphi} = c g M_0 \delta(R-r) \phi_0 \cos(\omega t)\,.
\end{equation}
This equation shows that the oscillating magnetization (\ref{deltaM}) is equivalent to a long  solenoid with an infinitely thin wire containing alternating current. This current creates electric and magnetic fields both inside and outside the magnet. 

We will look for a solution of Eqs.\,(\ref{MaxwellB1}) within the ansatz\footnote{The components of these vectors are given in cylindrical coordinates, e.g., $\vec E=(E_r,E_\varphi,E_z)$.}
$\delta\vec B = (0,0,B(r)\cos(\omega t))$, $\vec E=(0,E(r)\sin(\omega t),0)$ that corresponds to standing waves in the case $\omega R/c<1$. The amplitudes of these waves $B(r)$ and $E(r)$ are found analytically in terms of the Bessel functions $J_n$ and $Y_n$:
\begin{subequations}
\label{BE}
\begin{align}
    B(r) =& {\rm Re}[
    \kappa\theta(R-r) \sqrt{\varepsilon} J_0(\sqrt{\varepsilon}\omega r/c)Y_1(\omega R/c)\nonumber
    \\&
    +\kappa\theta(r-R)Y_0(\omega r/c)J_1(\sqrt{\varepsilon}\omega R/c)]\,,
    \label{B}\\
    E(r)=&{\rm Re}[
    \kappa\theta(R-r)J_1(\sqrt{\varepsilon}\omega r/c)Y_1(\omega R/c)\nonumber
    \\&
    +\kappa\theta(r-R)Y_1(\omega r/c)J_1(\sqrt{\varepsilon}\omega R/c)
    ]\,,
    \label{E}
\end{align}
\end{subequations}
where
\begin{equation}
    \kappa = \frac{-4\pi g M_0 \phi_0}{
    Y_0(\omega R/c)J_1(\sqrt{\varepsilon}\omega R/c)
    -\sqrt{\varepsilon}J_0(\sqrt{\varepsilon}\omega R/c)Y_1(\omega R/c)
    }\,.
    \label{kappa}
\end{equation}
is the normalization constant. Here
\begin{equation}
    \varepsilon = \epsilon+\frac{4\pi i \sigma}{\omega}
    \label{eps-complex}
\end{equation}
is the complex dielectric constant with its real part coinciding with the relative permittivity $\epsilon$ and with its imaginary part containing the electric conductivity $\sigma$. Thus, the solution (\ref{BE}) allows us to investigate dependence of the dark-matter-induced magnetic and electric fields on the electric conductivity $\sigma$.

The explicit solution (\ref{B}) allows us to find the magnetic field flux through the magnet cross section $S$:
\begin{align}
    \Phi &= \int_S \vec B\cdot d\vec s\nonumber\\
    &=2\pi\frac{Rc}{\omega} \cos(\omega t)
    {\rm Re}[\kappa J_1(\sqrt{\varepsilon}\omega R/c)Y_1(\omega R/c)]\,.
    \label{Phi}
\end{align}
This flux generates electromotive force (emf) in a pickup coil encircling the magnet that is found through the standard relation ${\cal E} = -d\Phi/dt$.

\subsection{Suppression of fields in conductors}
\label{SecSuppression}

An alternating magnetic field induces eddy currents on the surface of conductors. These currents partly shield the magnetic field inside the conductors and electric field on their surface. In this section, we estimate the suppression of the electric field on the boundary of an infinite cylindrical magnet. 

Let us consider a magnet with a high electric conductivity $\sigma$ such that
\begin{equation}
    \chi\equiv|\sqrt{\varepsilon}\omega R/c|\gg 1\,.
\end{equation}
In this regime, it is possible to apply the asymptotic expansion of the Bessel functions to show that $J_0(\sqrt{\varepsilon}\omega R/c)/J_1(\sqrt{\varepsilon}\omega R/c)\approx -i$.
As a result, we find that the electric field (\ref{E}) on the boundary is reduced to
\begin{equation}
    E(R)|_{\chi\gg1} \approx \frac{4\pi g}{2nc} \phi_0 M_0 \,,
    \label{ER}
\end{equation}
where $n={\rm Re}(\sqrt{\varepsilon})$ is the refractive index. Substitution of the complex dielectric constant $\varepsilon$ from Eq.  (\ref{eps-complex}) gives
\begin{equation}
    n=\sqrt{\frac{\epsilon + \sqrt{\epsilon^2+(4\pi\sigma/\omega)^2}}{2}}
    \approx\sqrt{\frac{2\pi\sigma}{\omega}}\equiv \frac c{\omega \delta}\,.
\end{equation}
Here 
\begin{equation}
    \delta = \frac c{\sqrt{2\pi\sigma\omega}}
\end{equation}
is the penetration depth. Comparing Eq.\,(\ref{ER}) with the electric field on the boundary for dielectrics ($\sigma=0$) we find the suppression of the electric field on the magnet boundary due to the eddy currents,
\begin{equation}
\xi=\frac{E(R)|_{\chi\gg1}}{E(R)|_{\sigma=0}}
=\frac1{2n}
\left(\sqrt{\epsilon}
\frac{J_0(\sqrt{\epsilon}\omega R/c)}{J_1(\sqrt{\epsilon}\omega R/c)}
-\frac{Y_0(\omega R/c)}{Y_1(\omega R/c)}
\right).
\label{xi}
\end{equation}
Note that the magnetic flux (\ref{Phi}), and, thus the induced emf, in a conductor is suppressed by the same factor (\ref{xi}) as compared with a dielectric with similar parameters. 

As an illustration, we compare the suppression (\ref{xi}) between neodymium and ferrite magnets. A typical neodymium magnet such as N52 has the electric conductivity $\sigma = 7.1\times 10^5$ S/m, see, e.g., \cite{N52}. Consider a cylindrical magnet of radius $R=5$\,cm and assume that the oscillation frequency is $f=1$\,MHz. For this magnet, we find that the suppression is sufficiently strong,
\begin{equation}
    \xi = 0.01\,.
\end{equation}

The electric conductivity of a typical ferrite magnet is low, $\sigma=10^{-4}$\,S/m (see, e.g., \cite{MagnetConductivity}). As a result, the imaginary part of the dielectric constant is very small, ${\rm Im(\sqrt{\varepsilon})}\ll1$, and the suppression of the electric and magnetic fields is negligible. The magnetic flux density for such a magnet is well approximated by Eq.\,(\ref{deltaB}). Thus, non-conducting magnets are more suitable for detection of the scalar field dark matter than their conducting alternatives.

\subsection{Enhancement of the magnetic flux in dielectrics}

Equation (\ref{kappa}) shows that the solution (\ref{BE}) may be singular for certain resonant frequencies. For dielectrics ($\sigma=0$), these frequencies correspond to the solutions of the transcendental equation
\begin{equation}
    Y_0(\omega R/c)J_1(\sqrt{\epsilon}\omega R/c)
    =\sqrt{\epsilon}J_0(\sqrt{\epsilon}\omega R/c)Y_1(\omega R/c)\,.
\label{resonance}
\end{equation}
This equation possesses non-trivial solutions for $\epsilon>1$. For physical applications, however, it is necessary to consider $\epsilon\gtrsim6$, to allow for the solutions satisfying $\omega R/c<1$.\footnote{We are considering the standing waves in the magnet which naturally appear when $\omega R/c<1$. For $\omega R/c>1$ an outgoing wave (radiation) should also be included.} 

In particular, for $\epsilon = 10$, Eq.\,(\ref{resonance}) possesses a solution $\omega R/c\approx 0.75$. Near this resonance frequency, the electric and magnetic fields inside the magnet may be strongly enhanced. Although this effect is not suitable for broadband detection, it may be exploited for studying a narrow band near some particular frequency. In practice, tuning to the resonance frequency may be difficult, as it would require an adjustment of the geometry of the magnet (for the cylindrical magnet, the radius $R$ is the only parameter).

In this paper, however, we do not consider resonance enhancement of the signal and focus rather on the broadband detection. Various scalar field dark matter detection experiments with resonance cavities were proposed in Ref.\,\cite{FMST}.


\section{Sensitivity estimates}
\label{SecSensitivity}

In this section, we estimate the sensitivity to the scalar field DM of two experimental setups which utilize the SQUID magnetometer and a low-noise rf amplifier, respectively. We consider also effects of a parasitic capacitance of a pickup solenoid and discuss the necessity of shielding from external rf noise.

\subsection{SQUID magnetometer}
\label{SecSensitivitySQUID}

The SQUID magnetometer is one of the most sensitive devices for measurements of magnetic fields and for searches of wave-like dark matter, see, e.g., \cite{DMradio,SHAFT}. In the SHAFT experiment \cite{SHAFT}, the record magnetic field sensitivity of the SQUID magnetometer of $\sqrt{S_B}=150$\,aT\,Hz$^{-1/2}$ was achieved. In this paper, we will assume the same noise level of the SQUID magnetometer in the searches of variations of fundamental constants with permanent magnets.

The signal is given by the flux of the magnetic field through the pickup loop of the magnetometer. This magnetic field is, in general, given by Eq.\,(\ref{B}), but for non-conducting magnets at low frequencies it reduces to 
\begin{equation}
B =  B_0 g\phi_0\cos(\omega t)\,,
\label{BB0}
\end{equation}
where $B_0=4\pi M_0$ is the static magnetic field inside a long cylindrical magnet. More precisely, in our estimates we will use the root-mean-square (rms) of this field, $\bar B = \frac1{\sqrt2} B_0 g\phi_0=g B_0 \frac{\sqrt{\rho_{\rm DM}}}{m_\phi}$. Thus, when the integration time exceeds the dark matter coherence time, $t>(\gamma f)^{-1}$, the signal-to-noise ratio is \cite{CASPEr}
\begin{equation}
    \text{SNR} = \frac{g B_0 \sqrt{\rho_{\rm DM}}}{m_\phi \sqrt{ S_B}}\left(
    \frac{t}{\gamma f}
    \right)^{1/4}\,,
    \label{SNR0}
\end{equation}
where $\gamma = 10^{-6}$ is the scalar field frequency bandwidth in the standard dark matter halo model. 

Equating SNR=1, we find the sensitivity of the experiment searching for the oscillations of the magnetic field of the permanent magnet with the SQUID magnetometer. 

Let us consider a non-conducting magnet with  magnetic field $B_0=0.3$\,T, and assume the local dark matter density $\rho_{\rm DM} = 0.4\,\text{GeV/}\text{cm}^3$. The integration time is taken $t=30$ days. For shorter measurement time, if $t<(\gamma f)^{-1}$, the last factor in Eq.\,(\ref{SNR0}) should be replaced as follows: $\left( \frac{t}{\gamma f} \right)^{1/4}\to \sqrt{t}$. See, e.g., Ref.\,\cite{CASPEr} for details.
    
\begin{figure*}
\begin{tabular}{cc}
    \includegraphics[width=8.8 cm]{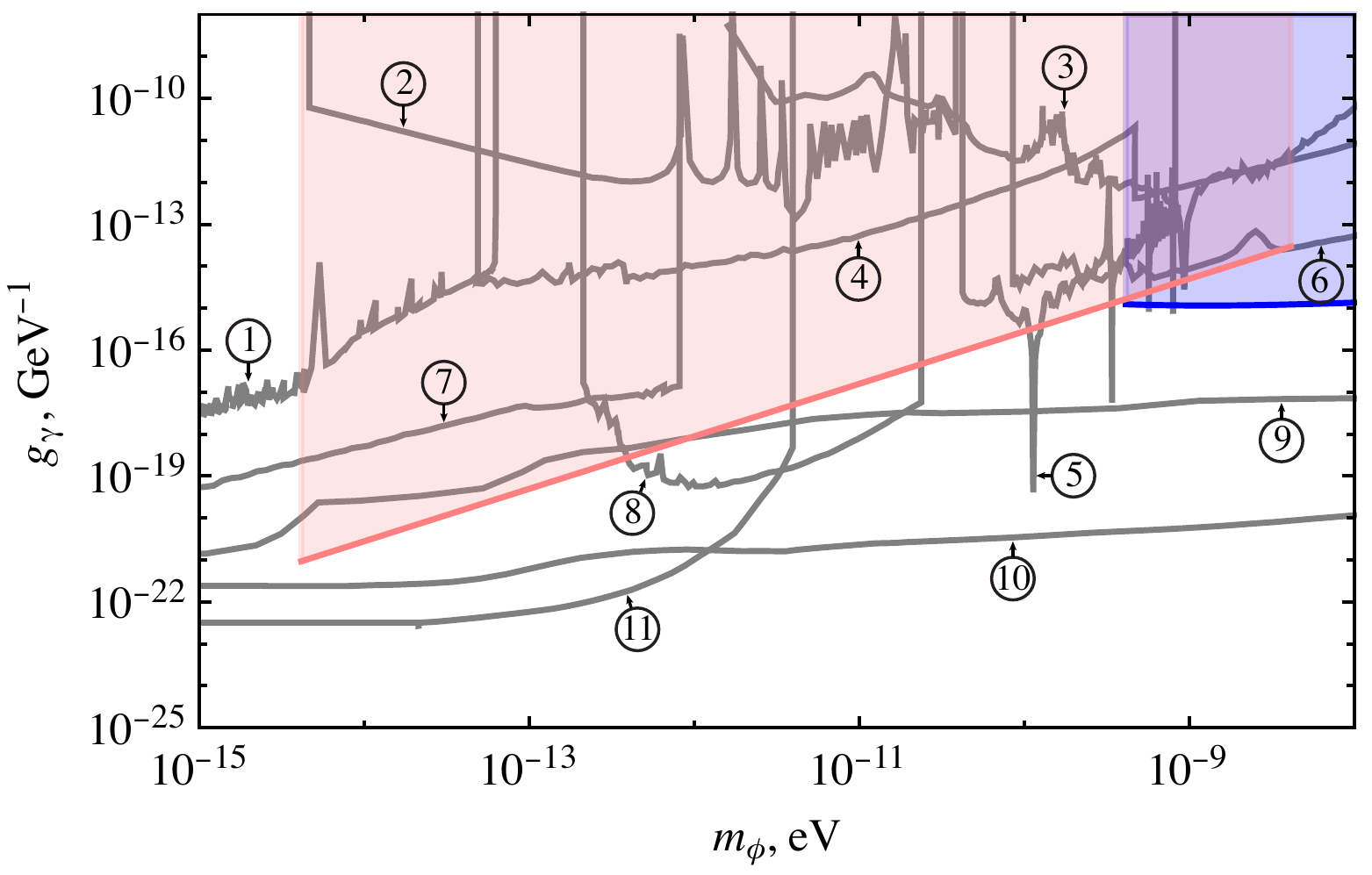}
    &
        \includegraphics[width=8.8 cm]{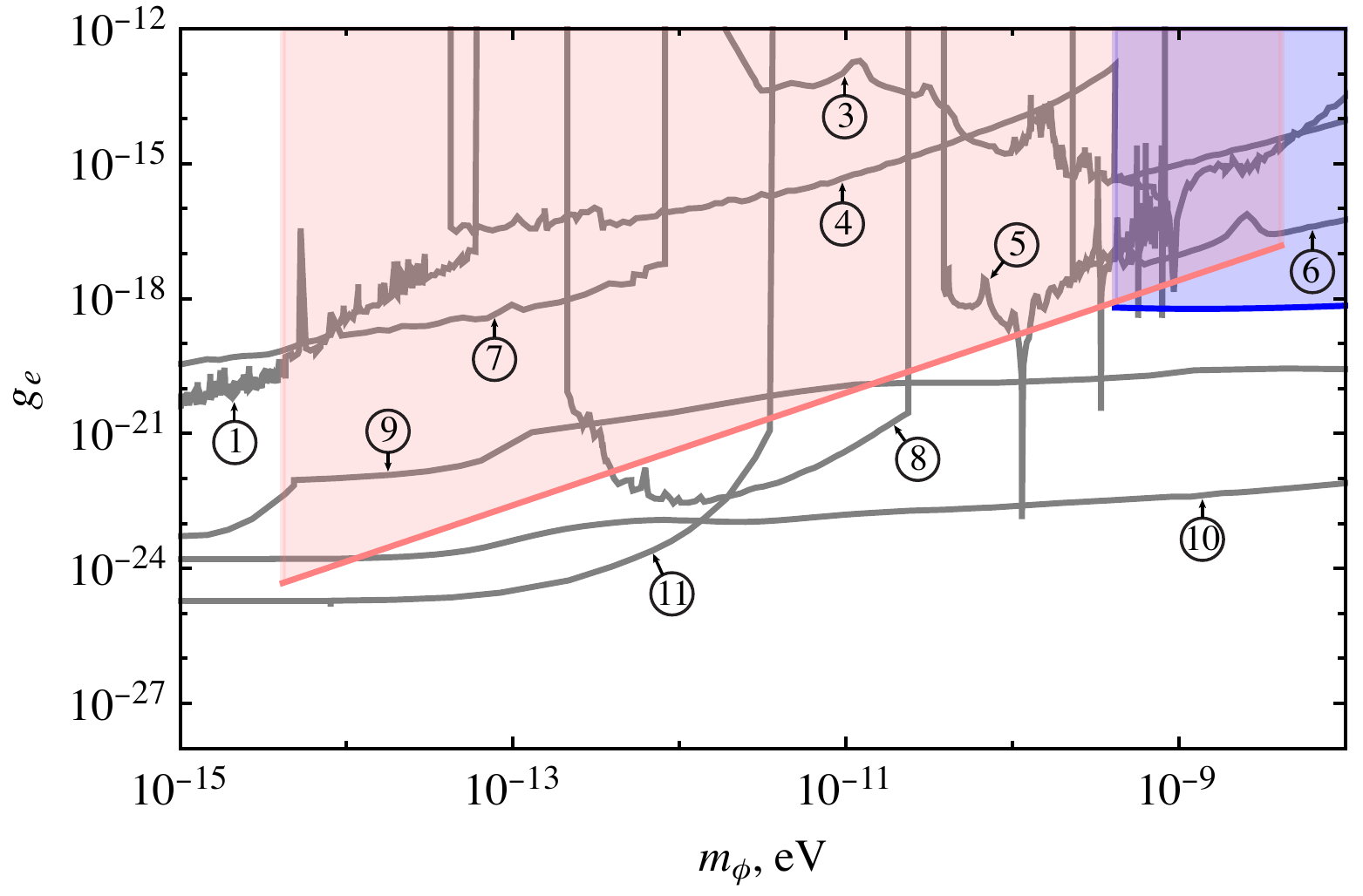}
\end{tabular}
    \caption{Projected sensitivity to the scalar-photon coupling $g_\gamma$ (left panel) and the scalar-electron coupling $g_e$ (right panel) of the experiment based on the oscillation of the magnetic field in a permanent magnet. Pink exclusion area represents the sensitivity of the experiment with the SQUID magnetometer, while the blue exclusion area corresponds to a low-noise rf amplifier. The sensitivity is compared with the results of other experiments: 1.\ H/Quartz/Sapphire \cite{HQuartzSapphire}; 2.\ Dynamic Decoupling \cite{DynamicDecoupling}; 3.\ Holometer \cite{Holometer}; 4.\ I$_2$ \cite{I2}; 5.\ DAMNED \cite{Damned}; 6.\ Cs-Cav \cite{CsCav}; 7.\ Dy \cite{DyQuartz}; 8.\ GEO600 \cite{GEO600}; 9.\ Fifth-Force \cite{FifthForce}; 10.\ E\"ot-Wash (EP) \cite{EotWash}; 11.\ MICROSCOPE \cite{Microscope}. The plot data for these experiments is taken from Ref.\,\cite{AxionLimits}.  }
    \label{FigSensitivitySQUID}
\end{figure*}

Equation (\ref{SNR0}) allows us to find limits on the combination of constant $g_\gamma$ and $g_e$ (\ref{geff}). Each of these constants may be constrained independently assuming that this constant gives the leading contribution. In Fig.\,\ref{FigSensitivitySQUID}, we present the expected exclusion plots for these constants if no signal is detected in this experiment.


\subsection{Low-noise RF amplifier}
\label{SecSensitivityRF}

For detection of the scalar-field dark matter, a boron ferrite permanent magnet may be used. Typical magnetic field of such a magnet is of order $B_0 = 0.3$\,T. Let us consider a cylindrical magnet with radius $R=5$\,cm inside a solenoid with $N=1000$ turns in the coil. The electric conductivity of such a magnet is low, so it may be considered as a dielectric. The dielectric constant is a function of unknown signal frequency. However, the value of the relative permittivity is bounded by the static dielectric constant $\epsilon\lesssim10$ (see, for example, \cite{BariumFerriteMagnet}), and the electric conductivity is very small, $\sigma\sim 10^{-4}$\, S/m \cite{MagnetConductivity}. As is shown in sec.\,\ref{SecSuppression}, for these values of the dielectric constant and electric conductivity and for frequencies below a few MHz, the suppression of the oscillating electric and magnetic fields in the magnet (\ref{BE}) is negligible, and the oscillating component of the magnetic field is given by Eq.\,(\ref{BB0}). 

For frequencies below the resonance frequency of the pickup solenoid the signal is formed by the emf in the coil
\begin{equation}
    {\cal E} = -N\frac{d\Phi}{dt}\,,
    \label{emf}
\end{equation}
where $\Phi$ is the magnetic field flux (\ref{Phi}). For the field (\ref{BB0}), the rms signal becomes 
\begin{equation}
    \bar{\cal E} = \pi R^2 N g B_0 \sqrt{\rho_{\rm DM}}\,.
    \label{signal}
\end{equation}
Apparently, this signal may be enhanced by using a larger and stronger magnet and by increasing the number of turns in the coil.

We assume that the permanent magnet with the coil may be cooled down to sufficiently low temperature such that the thermal magnetization noise becomes unimportant, see, for example, \cite{MagnetizationNoise}. In this case, the noise is determined by the noise level of the detector. In our estimates, we consider a commercial low-noise RF-amplifier HFC\,50\,D/E \cite{amplifier} with spectral noise floor $\sqrt{S_V}\geq 0.2\,\text{nV}/\sqrt{\text{Hz}}$ below 50\,MHz. Numerically, the spectral noise of this amplifier (in V/$\sqrt{\text{Hz}}$) may be modelled by the function \cite{FMST}
\begin{equation}
    \sqrt{S_V} = \sqrt{\frac{7.4185\times 10^{-14}}{f^{1.12}}+\frac{9.252\times 10^{-19}}{f^{0.176}}}\,,
\end{equation}
with $f$ being the signal frequency measured in hertz. Given this noise floor and the signal in Eq.\,(\ref{signal}), we find the signal-to-noise ratio
\begin{equation}
    \text{SNR} = \frac{\pi R^2 N g B_0 \sqrt{\rho_{\rm DM}}}{\sqrt{S_V}}
    \left(
    \frac{t}{\gamma f}
    \right)^{1/4}
    \,.
    \label{SNR1}
\end{equation}
Here we assume that the virialized dark matter has a bandwidth $\delta f = \gamma f$, where $\gamma \approx 10^{-6}$ in the standard dark matter halo model. The integration time $t$ is supposed to be greater than the scalar field coherence time, $t> (\gamma f)^{-1}$. For shorter measurement time, if $t<(\gamma f)^{-1}$, the last factor in Eq.\,(\ref{SNR1}) should be replaced as follows: $\left(
    \frac{t}{\gamma f}
    \right)^{1/4}\to \sqrt{t}$.

Equating SNR=1, we determine the sensitivity of the proposed magnet-based experiment, see Fig.\,\ref{FigSensitivitySQUID}. In principle, this sensitivity may be improved by using a stronger and larger magnet, by making a solenoid with larger number of turns in the coil, and by using a more sensitive detector. However, as is shown in the next section, these parameters are correlated and depend on the frequency band under investigation.

\subsection{Effect of parasitic capacitance of the pickup solenoid}
\label{SecParasiticCapacitance}

In the previous subsection, we assumed that the oscillating magnetic field could be detected with the use of a pickup solenoid. Naively, it seems advantageous to use a large solenoid with many turns in the coil. However, the efficiency of such a solenoid may be different for different frequencies. Any solenoid possesses a parasitic capacitance $C$ and, as a consequence, the principle resonance frequency
\begin{equation}
    f_0 = \frac1{2\pi\sqrt{LC}}\,,
    \label{f0}
\end{equation}
with $L$ being the inductance of the solenoid. The resonance frequency is an important characteristics of the solenoid, because for alternating currents with the frequencies above the resonance frequency it behaves rather as a capacitor, and Eq.\,(\ref{emf}) does not hold. As a result, only the frequencies $f<f_0$ may be efficiently  probed with the solenoid.

The inductance of a cylindrical solenoid with a ferrite core may be estimated as\footnote{In this subsection, we use SI units.} \cite{johnsonantenna}
\begin{equation}
    L=\mu_0\mu F_{l}N^2\frac{S}{l_{\rm coil}}\,,
    \label{Linductance}
\end{equation}
where $N$ is the number of turns in the coil, $S$ is the cross sectional area of the ferrite core, $l_{\rm coil}$ is the length of the coil, $\mu_0$ is the vacuum permeability, $\mu$ is relative permeability of the ferrite core, and $F_l$ is an empirical factor that depended on the ratio of the coil and core lengths. The latter is $F_l \approx 0.72$ when the core and the coil have the same length \cite{johnsonantenna}.
 
The capacitance of a single-layer coil with $N$ turns may be estimated as \cite{StrayCapacitance}
\begin{equation}
    C=\frac{C_0}{N-1}\,,
    \label{Cap}
\end{equation}
with $C_0$ the capacitance between two adjacent turns in the coil. Let $2r$ be the diameter of the conducting wire, $b$ be the width of the wire insulating layer with the relative permittivity $\epsilon$, and $p$ be the coil pitch. Then, the capacitance $C_0$ is \cite{StrayCapacitance}
\begin{equation}
C_0 = \frac{\pi^2 \epsilon_0 D}{\ln(F + \sqrt{F^2 - (1+b/r)^{2/\epsilon}})}\,,
\label{C0}
\end{equation}
where $D$ is the diameter of the coil and 
\begin{equation}
    F = \frac{p}{2r(1+b/r)^{1-\epsilon^{-1}}}\,.
\end{equation}

In particular, when the turns in the wire touch each other, $p=2(r+b)$, and when the wire insulating layer is thin, $b\ll r$, Eq.\,(\ref{C0}) simplifies:
\begin{equation}
    C_0 \approx \pi^2 \epsilon_0\epsilon D \frac r b\,.
\end{equation}
Substituting this expression into Eq.\,(\ref{Cap}), and assuming $N\gg1$, we find
\begin{equation}
    C = \frac{\pi^2\epsilon_0\epsilon r}{bN}\,.
    \label{C}
\end{equation}

Note that here we assume that the main contribution to the solenoid capacitance is given by the capacitance for each pair of adjacent turns. In general, there are also contributions from turn-to-shield capacitance and layer-to-layer capacitance for multilayer coils, see Ref.\,\cite{StrayCapacitance} for discussions. Here we assume that these contributions are subleading and ignore them.

Substituting the inductance (\ref{Linductance}) and capacitance (\ref{C}) into Eq.\,(\ref{f0}) we find the resonance frequency of the solenoid
\begin{equation}
f_0 = \frac1{2\pi} \sqrt{\frac{8b}{\pi^3 \epsilon_0\epsilon\mu_0\mu F_l D^3}}\,.
\label{f0result}
\end{equation}
Notably, the resonance frequency is independent of the number of turns in the coil. The main free parameter in this expression is the diameter of the coil $D$. 

For numerical estimates, let us consider a solenoid with the diameter $D=10$\,cm and a copper wire with kapton insulation with $\epsilon=3.5 $ and $b=0.01$\,mm. Assuming also that $\mu\sim 10$ for a ferrite magnet, we find
\begin{equation}
    f_0 = 0.48 \,\text{MHz}\,.
\end{equation}
For a coil with $D=1$\,cm, the resonance frequency becomes $f_0 = 15$\,MHz.

We stress that Eq.\,(\ref{f0result}) describes the resonance frequency of an isolated solenoid while in a real rexperiment the solenoid is supposed to be connected to a low-noise amplifier. In this case, the input capacitance of the amplifier should be considered. In particular, in the proposed above amplifier HFC 50 D/E the input capacitance is $C_{\rm in} = 6\,\text{pF}$ \cite{amplifier}, which is much larger than the parasitic capacitance of the solenoid. In this case, the input capacitance $C_{\rm in}$ should be substituted in Eq.\,(\ref{f0}) in place of the parasitic capacitance of the solenoid $C$. As a result, the capacitance is fixed in this equation, and only the inductance $L$ may be varied to adjust the resonance frequency $f_0$. To keep $f_0$ sufficiently high, one has to lower the inductance $L$. This limits the number of turns $N$ in the coil.

In particular, the pickup-coil with $N=1000$ turns has the inductance of order $L\approx 60$\,mH.  The corresponding circuit resonance frequency (\ref{f0}) is 
\begin{equation}
f_0 \approx 300\,\mbox{kHz}\,.
\end{equation}
Measurements at frequencies higher than this resonance will need to make use of careful electronic design, or an amplifier
with lower effective input capacitance.

\subsection{Effect of the magnetic shielding}

 The experimental setup should be isolated from external rf signals by a superconducting or mu-metal shield. One may wonder whether the shield itself can affect the dark matter signal. This question is important, in particular, in the
experiments searching for spin-dependent interactions of DM particles \cite{DMshielding} and in the Dark Matter Radio experiment \cite{DMradio}. In these references, it was noted that physical magnetic fields may be induced inside the magnetic shield due to the interaction with the DM particles. This magnetic field was considered as an additional contribution to the dark matter signal.

In the case of the scalar field dark matter, an electromagnetic field could, in principle, be produced by absorption of a scalar and emission of a photon by atoms. However, the magnetic shield can  hardly  create any additional contributions to the signal. The difference with dark photon case is that both the dark photon and the physical photon are vector fields, so the generated photon may be directed along the dark photon field. This is not the case for scalar field. The only vector characterizing the scalar particle is its momentum $\vec k$ which is a T-odd vector. The magnetic field $\vec B$ is a  pseudovector, so it cannot be directed along polar vector $\vec k$ due to parity conservation. This is an important difference with the case of a pseudoscalar field $a$ where $\vec B \propto \vec k a$ is allowed. The electric field $\vec E$ is a T-even vector, so, in principle, the relation $\frac{d\vec E}{dt}  \propto  \vec k$ is not forbidden. However, such field, directed along momentum, even if it exists, can hardly generate effects we are looking for, such as emf in a circular pick-up coil. Note also that the momentum of the dark matter particle is small compared to energy, $k \sim 10^{-3} m_\phi c$. 

Thus, we do not need to consider possible effects of the magnetic shield on the scalar dark matter induced fields. In the proposed experiment, cancellation of external rf noise is the only important effect of the magnetic shield.


\section{Summary}
\label{SecSummary}

In this paper, we considered the dark matter model with dark matter particles described by a classical oscillating scalar field $\phi$. This field can have non-minimal interaction with photon and electron fields with coupling constants $g_\gamma$ and $g_e$, respectively, as in Eq.\,(\ref{L}). This interaction can produce effects resembling oscillations of the fine structure constant and the electron mass. Although these oscillations may be extremely small, they can have a cumulative effect in solids with permanent magnetization and may lead to observable oscillation of the magnetic field in a permanent magnet. We theoretically study this effect and propose an experiment that could potentially probe such oscillation in the range of frequencies from 1\,Hz to a few MHz.

We stress that the effects of variations of fundamental constants should not be interpreted literally; rather, they represent an effective approach allowing one to take into account the interaction with the background scalar field (\ref{L}). Starting from the Lagrangian (\ref{L}) we derive the oscillating contributions to the magnetic field of a permanent magnet focusing on a long cylindrical magnet. Using the explicit solutions for the oscillating electric and magnetic fields we show that they may be strongly suppressed by eddy currents if the magnet has a good electric conductivity which is the case for most of neodymium and samarium cobalt magnets. Therefore, we show that non-conducting magnets such as barium-ferrite are suitable for the experiment aiming to detect variations of fundamental constants. We note also that non-conducting magnets may, in principle, have resonant frequencies at which the oscillating fields are enhanced. Approaches of detection of the scalar field dark matter with cavity resonators are studied more systematically in Ref.\,\cite{FMST}.

The oscillating magnetic field of permanent magnet may be detected either with a SQUID magnetometer or with a pickup solenoid connected to a low-noise rf amplifier. We consider both these possibilities and compare the sensitivities of these setups with other experiments \cite{HQuartzSapphire,DynamicDecoupling,Holometer,I2,Damned,CsCav,DyQuartz,GEO600,FifthForce,EotWash,Microscope}. As shown in Fig.\,\ref{FigSensitivitySQUID}, the projected sensitivity of the setup with the SQUID magnetometer exceeds the sensitivity of the experiments \cite{HQuartzSapphire,DynamicDecoupling,Holometer,I2,Damned,CsCav,DyQuartz} for both coupling constants considered in this work. This experiment may give new and complementary constraints to these constants. The setup based on a pickup solenoid and a low-noise rf amplifier has lower sensitivity and may be competitive only with such experiments as \cite{Holometer,I2,CsCav,Damned}. Another important limitation of this setup is the resonance frequency of the pickup solenoid, above which it becomes inefficient because of parasitic capacitance. As shown in Sec.\,\ref{SecParasiticCapacitance}, this resonance frequency may vary from one to a few MHz, depending on the parameters of the solenoid.

In this paper, we considered only the linear coupling of the scalar field with the Standard Matter fields while quadratic couplings introduced in Refs.\,\cite{Stadnik2015,Stadnik2015a,Stadnik2016,Hees2018,Stadnik2019,Kim2022,Banerjee2022} are of interest as well. In this case, the field $\phi$ may be scalar or pseudoscalar (axion) field. We leave this question for future study.

\vspace{3mm}
\textit{Acknowledgements} --- 
The work of VVF and IBS was supported by the Australian Research Council Grants No.\ DP230101058 and DP200100150. The work of DB and OT is supported in part by the Cluster of Excellence ``Precision Physics, Fundamental Interactions, and Structure of Matter'' (PRISMA+ EXC 2118/1) funded by the German Research Foundation (DFG) within the German Excellence Strategy (Project ID 39083149). This article is based in part upon work from COST Action COSMIC WISPers CA21106, supported by COST (European Cooperation in Science and Technology). The work of AS is supported in part by the NSF CAREER grant PHY-2145162 and by the US Department of Energy, Office of High Energy Physics, under the QuantISED program, FWP 100667.



\begin{thebibliography}{37}%
\makeatletter
\providecommand \@ifxundefined [1]{%
 \@ifx{#1\undefined}
}%
\providecommand \@ifnum [1]{%
 \ifnum #1\expandafter \@firstoftwo
 \else \expandafter \@secondoftwo
 \fi
}%
\providecommand \@ifx [1]{%
 \ifx #1\expandafter \@firstoftwo
 \else \expandafter \@secondoftwo
 \fi
}%
\providecommand \natexlab [1]{#1}%
\providecommand \enquote  [1]{``#1''}%
\providecommand \bibnamefont  [1]{#1}%
\providecommand \bibfnamefont [1]{#1}%
\providecommand \citenamefont [1]{#1}%
\providecommand \href@noop [0]{\@secondoftwo}%
\providecommand \href [0]{\begingroup \@sanitize@url \@href}%
\providecommand \@href[1]{\@@startlink{#1}\@@href}%
\providecommand \@@href[1]{\endgroup#1\@@endlink}%
\providecommand \@sanitize@url [0]{\catcode `\\12\catcode `\$12\catcode
  `\&12\catcode `\#12\catcode `\^12\catcode `\_12\catcode `\%12\relax}%
\providecommand \@@startlink[1]{}%
\providecommand \@@endlink[0]{}%
\providecommand \url  [0]{\begingroup\@sanitize@url \@url }%
\providecommand \@url [1]{\endgroup\@href {#1}{\urlprefix }}%
\providecommand \urlprefix  [0]{URL }%
\providecommand \Eprint [0]{\href }%
\providecommand \doibase [0]{https://doi.org/}%
\providecommand \selectlanguage [0]{\@gobble}%
\providecommand \bibinfo  [0]{\@secondoftwo}%
\providecommand \bibfield  [0]{\@secondoftwo}%
\providecommand \translation [1]{[#1]}%
\providecommand \BibitemOpen [0]{}%
\providecommand \bibitemStop [0]{}%
\providecommand \bibitemNoStop [0]{.\EOS\space}%
\providecommand \EOS [0]{\spacefactor3000\relax}%
\providecommand \BibitemShut  [1]{\csname bibitem#1\endcsname}%
\let\auto@bib@innerbib\@empty
\bibitem [{\citenamefont {Bertone}(2010)}]{bertone2010}%
  \BibitemOpen
  \bibinfo {editor} {\bibfnamefont {G.}~\bibnamefont {Bertone}},\ ed.,\ \href
  {https://doi.org/10.1017/CBO9780511770739} {\emph {\bibinfo {title} {Particle
  Dark Matter: Observations, Models and Searches}}}\ (\bibinfo  {publisher}
  {Cambridge University Press},\ \bibinfo {address} {Cambridge},\ \bibinfo
  {year} {2010})\BibitemShut {NoStop}%
\bibitem [{\citenamefont {Kimball}\ and\ \citenamefont {van
  Bibber}(2022)}]{kimball2022search}%
  \BibitemOpen
  \bibfield  {author} {\bibinfo {author} {\bibfnamefont {D.~F.~J.}\
  \bibnamefont {Kimball}}\ and\ \bibinfo {author} {\bibfnamefont
  {K.}~\bibnamefont {van Bibber}},\ }\href@noop {} {\emph {\bibinfo {title}
  {The Search for Ultralight Bosonic Dark Matter}}}\ (\bibinfo  {publisher}
  {Springer},\ \bibinfo {year} {2022})\BibitemShut {NoStop}%
\bibitem [{\citenamefont {Ellis}\ \emph {et~al.}(1989)\citenamefont {Ellis},
  \citenamefont {Kalara}, \citenamefont {Olive},\ and\ \citenamefont
  {Wetterich}}]{oscillation}%
  \BibitemOpen
  \bibfield  {author} {\bibinfo {author} {\bibfnamefont {J.}~\bibnamefont
  {Ellis}}, \bibinfo {author} {\bibfnamefont {S.}~\bibnamefont {Kalara}},
  \bibinfo {author} {\bibfnamefont {K.}~\bibnamefont {Olive}},\ and\ \bibinfo
  {author} {\bibfnamefont {C.}~\bibnamefont {Wetterich}},\ }\href
  {https://doi.org/https://doi.org/10.1016/0370-2693(89)90669-2} {\bibfield
  {journal} {\bibinfo  {journal} {Phys. Lett. B}\ }\textbf {\bibinfo {volume}
  {228}},\ \bibinfo {pages} {264} (\bibinfo {year} {1989})}\BibitemShut
  {NoStop}%
\bibitem [{\citenamefont {Arvanitaki}\ \emph {et~al.}(2015)\citenamefont
  {Arvanitaki}, \citenamefont {Huang},\ and\ \citenamefont
  {Van~Tilburg}}]{Arvanitaki2015}%
  \BibitemOpen
  \bibfield  {author} {\bibinfo {author} {\bibfnamefont {A.}~\bibnamefont
  {Arvanitaki}}, \bibinfo {author} {\bibfnamefont {J.}~\bibnamefont {Huang}},\
  and\ \bibinfo {author} {\bibfnamefont {K.}~\bibnamefont {Van~Tilburg}},\
  }\href {https://doi.org/10.1103/PhysRevD.91.015015} {\bibfield  {journal}
  {\bibinfo  {journal} {Phys. Rev. D}\ }\textbf {\bibinfo {volume} {91}},\
  \bibinfo {pages} {015015} (\bibinfo {year} {2015})}\BibitemShut {NoStop}%
\bibitem [{\citenamefont {Antypas}\ \emph {et~al.}(2020)\citenamefont
  {Antypas}, \citenamefont {Budker}, \citenamefont {Flambaum}, \citenamefont
  {Kozlov}, \citenamefont {Perez},\ and\ \citenamefont {Ye}}]{Antypas2020}%
  \BibitemOpen
  \bibfield  {author} {\bibinfo {author} {\bibfnamefont {D.}~\bibnamefont
  {Antypas}}, \bibinfo {author} {\bibfnamefont {D.}~\bibnamefont {Budker}},
  \bibinfo {author} {\bibfnamefont {V.~V.}\ \bibnamefont {Flambaum}}, \bibinfo
  {author} {\bibfnamefont {M.~G.}\ \bibnamefont {Kozlov}}, \bibinfo {author}
  {\bibfnamefont {G.}~\bibnamefont {Perez}},\ and\ \bibinfo {author}
  {\bibfnamefont {J.}~\bibnamefont {Ye}},\ }\href
  {https://doi.org/10.1002/andp.201900566} {\bibfield  {journal} {\bibinfo
  {journal} {Ann. Phys.}\ }\textbf {\bibinfo {volume} {532}},\ \bibinfo {pages}
  {1900566} (\bibinfo {year} {2020})}\BibitemShut {NoStop}%
\bibitem [{\citenamefont {Antypas}\ \emph {et~al.}(2022)\citenamefont {Antypas}
  \emph {et~al.}}]{Snowmass}%
  \BibitemOpen
  \bibfield  {author} {\bibinfo {author} {\bibfnamefont {D.}~\bibnamefont
  {Antypas}} \emph {et~al.},\ }\href@noop {} {\bibinfo {title} {{New Horizons:
  Scalar and vector ultralight dark matter}}} (\bibinfo {year} {2022}),\
  \Eprint {https://arxiv.org/abs/2203.14915} {arXiv:2203.14915 [hep-ex]}
  \BibitemShut {NoStop}%
\bibitem [{1()}]{1}%
  \BibitemOpen
  \href@noop {} {}\bibinfo {note} {Such an experiment is currently in progress
  at Boston University.}\BibitemShut {Stop}%
\bibitem [{\citenamefont {Campbell}\ \emph {et~al.}(2021)\citenamefont
  {Campbell}, \citenamefont {McAllister}, \citenamefont {Goryachev},
  \citenamefont {Ivanov},\ and\ \citenamefont {Tobar}}]{HQuartzSapphire}%
  \BibitemOpen
  \bibfield  {author} {\bibinfo {author} {\bibfnamefont {W.~M.}\ \bibnamefont
  {Campbell}}, \bibinfo {author} {\bibfnamefont {B.~T.}\ \bibnamefont
  {McAllister}}, \bibinfo {author} {\bibfnamefont {M.}~\bibnamefont
  {Goryachev}}, \bibinfo {author} {\bibfnamefont {E.~N.}\ \bibnamefont
  {Ivanov}},\ and\ \bibinfo {author} {\bibfnamefont {M.~E.}\ \bibnamefont
  {Tobar}},\ }\href {https://doi.org/10.1103/PhysRevLett.126.071301} {\bibfield
   {journal} {\bibinfo  {journal} {Phys. Rev. Lett.}\ }\textbf {\bibinfo
  {volume} {126}},\ \bibinfo {pages} {071301} (\bibinfo {year}
  {2021})}\BibitemShut {NoStop}%
\bibitem [{\citenamefont {Aharony}\ \emph {et~al.}(2021)\citenamefont
  {Aharony}, \citenamefont {Akerman}, \citenamefont {Ozeri}, \citenamefont
  {Perez}, \citenamefont {Savoray},\ and\ \citenamefont
  {Shaniv}}]{DynamicDecoupling}%
  \BibitemOpen
  \bibfield  {author} {\bibinfo {author} {\bibfnamefont {S.}~\bibnamefont
  {Aharony}}, \bibinfo {author} {\bibfnamefont {N.}~\bibnamefont {Akerman}},
  \bibinfo {author} {\bibfnamefont {R.}~\bibnamefont {Ozeri}}, \bibinfo
  {author} {\bibfnamefont {G.}~\bibnamefont {Perez}}, \bibinfo {author}
  {\bibfnamefont {I.}~\bibnamefont {Savoray}},\ and\ \bibinfo {author}
  {\bibfnamefont {R.}~\bibnamefont {Shaniv}},\ }\href
  {https://doi.org/10.1103/PhysRevD.103.075017} {\bibfield  {journal} {\bibinfo
   {journal} {Phys. Rev. D}\ }\textbf {\bibinfo {volume} {103}},\ \bibinfo
  {pages} {075017} (\bibinfo {year} {2021})}\BibitemShut {NoStop}%
\bibitem [{\citenamefont {Aiello}\ \emph {et~al.}(2022)\citenamefont {Aiello},
  \citenamefont {Richardson}, \citenamefont {Vermeulen}, \citenamefont {Grote},
  \citenamefont {Hogan}, \citenamefont {Kwon},\ and\ \citenamefont
  {Stoughton}}]{Holometer}%
  \BibitemOpen
  \bibfield  {author} {\bibinfo {author} {\bibfnamefont {L.}~\bibnamefont
  {Aiello}}, \bibinfo {author} {\bibfnamefont {J.~W.}\ \bibnamefont
  {Richardson}}, \bibinfo {author} {\bibfnamefont {S.~M.}\ \bibnamefont
  {Vermeulen}}, \bibinfo {author} {\bibfnamefont {H.}~\bibnamefont {Grote}},
  \bibinfo {author} {\bibfnamefont {C.}~\bibnamefont {Hogan}}, \bibinfo
  {author} {\bibfnamefont {O.}~\bibnamefont {Kwon}},\ and\ \bibinfo {author}
  {\bibfnamefont {C.}~\bibnamefont {Stoughton}},\ }\href
  {https://doi.org/10.1103/PhysRevLett.128.121101} {\bibfield  {journal}
  {\bibinfo  {journal} {Phys. Rev. Lett.}\ }\textbf {\bibinfo {volume} {128}},\
  \bibinfo {pages} {121101} (\bibinfo {year} {2022})}\BibitemShut {NoStop}%
\bibitem [{\citenamefont {Oswald}\ \emph {et~al.}(2022)\citenamefont {Oswald}
  \emph {et~al.}}]{I2}%
  \BibitemOpen
  \bibfield  {author} {\bibinfo {author} {\bibfnamefont {R.}~\bibnamefont
  {Oswald}} \emph {et~al.},\ }\href
  {https://doi.org/10.1103/PhysRevLett.129.031302} {\bibfield  {journal}
  {\bibinfo  {journal} {Phys. Rev. Lett.}\ }\textbf {\bibinfo {volume} {129}},\
  \bibinfo {pages} {031302} (\bibinfo {year} {2022})}\BibitemShut {NoStop}%
\bibitem [{\citenamefont {Savalle}\ \emph {et~al.}(2021)\citenamefont {Savalle}
  \emph {et~al.}}]{Damned}%
  \BibitemOpen
  \bibfield  {author} {\bibinfo {author} {\bibfnamefont {E.}~\bibnamefont
  {Savalle}} \emph {et~al.},\ }\href
  {https://doi.org/10.1103/PhysRevLett.126.051301} {\bibfield  {journal}
  {\bibinfo  {journal} {Phys. Rev. Lett.}\ }\textbf {\bibinfo {volume} {126}},\
  \bibinfo {pages} {051301} (\bibinfo {year} {2021})}\BibitemShut {NoStop}%
\bibitem [{\citenamefont {Tretiak}\ \emph {et~al.}(2022)\citenamefont {Tretiak}
  \emph {et~al.}}]{CsCav}%
  \BibitemOpen
  \bibfield  {author} {\bibinfo {author} {\bibfnamefont {O.}~\bibnamefont
  {Tretiak}} \emph {et~al.},\ }\href
  {https://doi.org/10.1103/PhysRevLett.129.031301} {\bibfield  {journal}
  {\bibinfo  {journal} {Phys. Rev. Lett.}\ }\textbf {\bibinfo {volume} {129}},\
  \bibinfo {pages} {031301} (\bibinfo {year} {2022})}\BibitemShut {NoStop}%
\bibitem [{\citenamefont {Zhang}\ \emph {et~al.}(2022)\citenamefont {Zhang}
  \emph {et~al.}}]{DyQuartz}%
  \BibitemOpen
  \bibfield  {author} {\bibinfo {author} {\bibfnamefont {X.}~\bibnamefont
  {Zhang}} \emph {et~al.},\ }\href@noop {} {\bibinfo {title} {{Search for
  ultralight dark matter with spectroscopy of radio-frequency atomic
  transitions}}} (\bibinfo {year} {2022}),\ \Eprint
  {https://arxiv.org/abs/2212.04413} {arXiv:2212.04413 [physics.atom-ph]}
  \BibitemShut {NoStop}%
\bibitem [{\citenamefont {Vermeulen}\ \emph {et~al.}(2021)\citenamefont
  {Vermeulen} \emph {et~al.}}]{GEO600}%
  \BibitemOpen
  \bibfield  {author} {\bibinfo {author} {\bibfnamefont {S.}~\bibnamefont
  {Vermeulen}} \emph {et~al.},\ }\href
  {https://doi.org/https://doi.org/10.1038/s41586-021-04031-y} {\bibfield
  {journal} {\bibinfo  {journal} {Nature}\ }\textbf {\bibinfo {volume} {600}},\
  \bibinfo {pages} {424–428} (\bibinfo {year} {2021})}\BibitemShut {NoStop}%
\bibitem [{\citenamefont {Arvanitaki}\ \emph {et~al.}(2016)\citenamefont
  {Arvanitaki}, \citenamefont {Dimopoulos},\ and\ \citenamefont
  {Van~Tilburg}}]{FifthForce}%
  \BibitemOpen
  \bibfield  {author} {\bibinfo {author} {\bibfnamefont {A.}~\bibnamefont
  {Arvanitaki}}, \bibinfo {author} {\bibfnamefont {S.}~\bibnamefont
  {Dimopoulos}},\ and\ \bibinfo {author} {\bibfnamefont {K.}~\bibnamefont
  {Van~Tilburg}},\ }\href {https://doi.org/10.1103/PhysRevLett.116.031102}
  {\bibfield  {journal} {\bibinfo  {journal} {Phys. Rev. Lett.}\ }\textbf
  {\bibinfo {volume} {116}},\ \bibinfo {pages} {031102} (\bibinfo {year}
  {2016})}\BibitemShut {NoStop}%
\bibitem [{\citenamefont {Shaw}\ \emph {et~al.}(2022)\citenamefont {Shaw},
  \citenamefont {Ross}, \citenamefont {Hagedorn}, \citenamefont {Adelberger},\
  and\ \citenamefont {Gundlach}}]{EotWash}%
  \BibitemOpen
  \bibfield  {author} {\bibinfo {author} {\bibfnamefont {E.~A.}\ \bibnamefont
  {Shaw}}, \bibinfo {author} {\bibfnamefont {M.~P.}\ \bibnamefont {Ross}},
  \bibinfo {author} {\bibfnamefont {C.~A.}\ \bibnamefont {Hagedorn}}, \bibinfo
  {author} {\bibfnamefont {E.~G.}\ \bibnamefont {Adelberger}},\ and\ \bibinfo
  {author} {\bibfnamefont {J.~H.}\ \bibnamefont {Gundlach}},\ }\href
  {https://doi.org/10.1103/PhysRevD.105.042007} {\bibfield  {journal} {\bibinfo
   {journal} {Phys. Rev. D}\ }\textbf {\bibinfo {volume} {105}},\ \bibinfo
  {pages} {042007} (\bibinfo {year} {2022})}\BibitemShut {NoStop}%
\bibitem [{\citenamefont {Berg\'e}\ \emph {et~al.}(2018)\citenamefont
  {Berg\'e}, \citenamefont {Brax}, \citenamefont {M\'etris}, \citenamefont
  {Pernot-Borr\`as}, \citenamefont {Touboul},\ and\ \citenamefont
  {Uzan}}]{Microscope}%
  \BibitemOpen
  \bibfield  {author} {\bibinfo {author} {\bibfnamefont {J.}~\bibnamefont
  {Berg\'e}}, \bibinfo {author} {\bibfnamefont {P.}~\bibnamefont {Brax}},
  \bibinfo {author} {\bibfnamefont {G.}~\bibnamefont {M\'etris}}, \bibinfo
  {author} {\bibfnamefont {M.}~\bibnamefont {Pernot-Borr\`as}}, \bibinfo
  {author} {\bibfnamefont {P.}~\bibnamefont {Touboul}},\ and\ \bibinfo {author}
  {\bibfnamefont {J.-P.}\ \bibnamefont {Uzan}},\ }\href
  {https://doi.org/10.1103/PhysRevLett.120.141101} {\bibfield  {journal}
  {\bibinfo  {journal} {Phys. Rev. Lett.}\ }\textbf {\bibinfo {volume} {120}},\
  \bibinfo {pages} {141101} (\bibinfo {year} {2018})}\BibitemShut {NoStop}%
\bibitem [{N52()}]{N52}%
  \BibitemOpen
  \href@noop {} {\bibinfo {title} {{Nd-Fe-B} magnet: Demagnetization curves at
  elevated temperature}},\ \bibinfo {howpublished}
  {\url{https://www.shinetsu.co.jp/serem/e/download/N52sheet.pdf}},\ \bibinfo
  {note} {accessed: 2023-01-02.}\BibitemShut {Stop}%
\bibitem [{\citenamefont {Marouani}\ \emph {et~al.}(2021)\citenamefont
  {Marouani} \emph {et~al.}}]{MagnetConductivity}%
  \BibitemOpen
  \bibfield  {author} {\bibinfo {author} {\bibfnamefont {Y.}~\bibnamefont
  {Marouani}} \emph {et~al.},\ }\href {https://doi.org/10.1039/D0RA09465J}
  {\bibfield  {journal} {\bibinfo  {journal} {RSC Adv.}\ }\textbf {\bibinfo
  {volume} {11}},\ \bibinfo {pages} {1531} (\bibinfo {year}
  {2021})}\BibitemShut {NoStop}%
\bibitem [{\citenamefont {Flambaum}\ \emph {et~al.}(2022)\citenamefont
  {Flambaum}, \citenamefont {McAllister}, \citenamefont {Samsonov},\ and\
  \citenamefont {Tobar}}]{FMST}%
  \BibitemOpen
  \bibfield  {author} {\bibinfo {author} {\bibfnamefont {V.~V.}\ \bibnamefont
  {Flambaum}}, \bibinfo {author} {\bibfnamefont {B.~T.}\ \bibnamefont
  {McAllister}}, \bibinfo {author} {\bibfnamefont {I.~B.}\ \bibnamefont
  {Samsonov}},\ and\ \bibinfo {author} {\bibfnamefont {M.~E.}\ \bibnamefont
  {Tobar}},\ }\href {https://doi.org/10.1103/PhysRevD.106.055037} {\bibfield
  {journal} {\bibinfo  {journal} {Phys. Rev. D}\ }\textbf {\bibinfo {volume}
  {106}},\ \bibinfo {pages} {055037} (\bibinfo {year} {2022})}\BibitemShut
  {NoStop}%
\bibitem [{\citenamefont {Chaudhuri}\ \emph {et~al.}(2015)\citenamefont
  {Chaudhuri}, \citenamefont {Graham}, \citenamefont {Irwin}, \citenamefont
  {Mardon}, \citenamefont {Rajendran},\ and\ \citenamefont {Zhao}}]{DMradio}%
  \BibitemOpen
  \bibfield  {author} {\bibinfo {author} {\bibfnamefont {S.}~\bibnamefont
  {Chaudhuri}}, \bibinfo {author} {\bibfnamefont {P.~W.}\ \bibnamefont
  {Graham}}, \bibinfo {author} {\bibfnamefont {K.}~\bibnamefont {Irwin}},
  \bibinfo {author} {\bibfnamefont {J.}~\bibnamefont {Mardon}}, \bibinfo
  {author} {\bibfnamefont {S.}~\bibnamefont {Rajendran}},\ and\ \bibinfo
  {author} {\bibfnamefont {Y.}~\bibnamefont {Zhao}},\ }\href
  {https://doi.org/10.1103/PhysRevD.92.075012} {\bibfield  {journal} {\bibinfo
  {journal} {Phys. Rev. D}\ }\textbf {\bibinfo {volume} {92}},\ \bibinfo
  {pages} {075012} (\bibinfo {year} {2015})}\BibitemShut {NoStop}%
\bibitem [{\citenamefont {Gramolin}\ \emph {et~al.}(2021)\citenamefont
  {Gramolin}, \citenamefont {Aybas}, \citenamefont {Johnson}, \citenamefont
  {Adam},\ and\ \citenamefont {Sushkov}}]{SHAFT}%
  \BibitemOpen
  \bibfield  {author} {\bibinfo {author} {\bibfnamefont {A.~V.}\ \bibnamefont
  {Gramolin}}, \bibinfo {author} {\bibfnamefont {D.}~\bibnamefont {Aybas}},
  \bibinfo {author} {\bibfnamefont {D.}~\bibnamefont {Johnson}}, \bibinfo
  {author} {\bibfnamefont {J.}~\bibnamefont {Adam}},\ and\ \bibinfo {author}
  {\bibfnamefont {A.~O.}\ \bibnamefont {Sushkov}},\ }\href
  {https://doi.org/10.1038/s41567-020-1006-6} {\bibfield  {journal} {\bibinfo
  {journal} {Nature Physics}\ }\textbf {\bibinfo {volume} {17}},\ \bibinfo
  {pages} {79–84} (\bibinfo {year} {2021})}\BibitemShut {NoStop}%
\bibitem [{\citenamefont {Budker}\ \emph {et~al.}(2014)\citenamefont {Budker},
  \citenamefont {Graham}, \citenamefont {Ledbetter}, \citenamefont
  {Rajendran},\ and\ \citenamefont {Sushkov}}]{CASPEr}%
  \BibitemOpen
  \bibfield  {author} {\bibinfo {author} {\bibfnamefont {D.}~\bibnamefont
  {Budker}}, \bibinfo {author} {\bibfnamefont {P.~W.}\ \bibnamefont {Graham}},
  \bibinfo {author} {\bibfnamefont {M.}~\bibnamefont {Ledbetter}}, \bibinfo
  {author} {\bibfnamefont {S.}~\bibnamefont {Rajendran}},\ and\ \bibinfo
  {author} {\bibfnamefont {A.~O.}\ \bibnamefont {Sushkov}},\ }\href
  {https://doi.org/10.1103/PhysRevX.4.021030} {\bibfield  {journal} {\bibinfo
  {journal} {Phys. Rev. X}\ }\textbf {\bibinfo {volume} {4}},\ \bibinfo {pages}
  {021030} (\bibinfo {year} {2014})}\BibitemShut {NoStop}%
\bibitem [{\citenamefont {O'Hare}(2020)}]{AxionLimits}%
  \BibitemOpen
  \bibfield  {author} {\bibinfo {author} {\bibfnamefont {C.}~\bibnamefont
  {O'Hare}},\ }\href {https://doi.org/10.5281/zenodo.3932430} {\bibinfo {title}
  {cajohare/axionlimits: Axionlimits}},\ \bibinfo {howpublished}
  {\url{https://cajohare.github.io/AxionLimits/}} (\bibinfo {year}
  {2020})\BibitemShut {NoStop}%
\bibitem [{\citenamefont {Sholiyi}\ \emph {et~al.}(2014)\citenamefont
  {Sholiyi}, \citenamefont {Lee},\ and\ \citenamefont
  {Williams}}]{BariumFerriteMagnet}%
  \BibitemOpen
  \bibfield  {author} {\bibinfo {author} {\bibfnamefont {O.}~\bibnamefont
  {Sholiyi}}, \bibinfo {author} {\bibfnamefont {J.}~\bibnamefont {Lee}},\ and\
  \bibinfo {author} {\bibfnamefont {J.~D.}\ \bibnamefont {Williams}},\ }\href
  {https://doi.org/10.1063/1.4891936} {\bibfield  {journal} {\bibinfo
  {journal} {AIP Advances}\ }\textbf {\bibinfo {volume} {4}},\ \bibinfo {pages}
  {077136} (\bibinfo {year} {2014})}\BibitemShut {NoStop}%
\bibitem [{\citenamefont {Eckel}\ \emph {et~al.}(2009)\citenamefont {Eckel},
  \citenamefont {Sushkov},\ and\ \citenamefont
  {Lamoreaux}}]{MagnetizationNoise}%
  \BibitemOpen
  \bibfield  {author} {\bibinfo {author} {\bibfnamefont {S.}~\bibnamefont
  {Eckel}}, \bibinfo {author} {\bibfnamefont {A.~O.}\ \bibnamefont {Sushkov}},\
  and\ \bibinfo {author} {\bibfnamefont {S.~K.}\ \bibnamefont {Lamoreaux}},\
  }\href {https://doi.org/10.1103/PhysRevB.79.014422} {\bibfield  {journal}
  {\bibinfo  {journal} {Phys. Rev. B}\ }\textbf {\bibinfo {volume} {79}},\
  \bibinfo {pages} {014422} (\bibinfo {year} {2009})}\BibitemShut {NoStop}%
\bibitem [{amp()}]{amplifier}%
  \BibitemOpen
  \href@noop {} {\bibinfo {title} {{HFC 50 D/E}, dual cryogenic ultra low noise
  rf-amplifier}},\ \bibinfo {howpublished}
  {\url{https://www.stahl-electronics.com/bilder/Datasheet_HFC50DE_2-38.pdf}},\
  \bibinfo {note} {accessed: 2023-01-20.}\BibitemShut {Stop}%
\bibitem [{\citenamefont {Johnson}(1993)}]{johnsonantenna}%
  \BibitemOpen
  \bibfield  {author} {\bibinfo {author} {\bibfnamefont {R.~C.}\ \bibnamefont
  {Johnson}},\ }\href@noop {} {\emph {\bibinfo {title} {Antenna Engineering
  Handbook}}}\ (\bibinfo  {publisher} {McGraw-Hill Professional},\ \bibinfo
  {year} {1993})\BibitemShut {NoStop}%
\bibitem [{\citenamefont {Grandi}\ \emph {et~al.}(1999)\citenamefont {Grandi},
  \citenamefont {Kazimierczuk}, \citenamefont {Massarini},\ and\ \citenamefont
  {Reggiani}}]{StrayCapacitance}%
  \BibitemOpen
  \bibfield  {author} {\bibinfo {author} {\bibfnamefont {G.}~\bibnamefont
  {Grandi}}, \bibinfo {author} {\bibfnamefont {M.}~\bibnamefont
  {Kazimierczuk}}, \bibinfo {author} {\bibfnamefont {A.}~\bibnamefont
  {Massarini}},\ and\ \bibinfo {author} {\bibfnamefont {U.}~\bibnamefont
  {Reggiani}},\ }\href {https://doi.org/10.1109/28.793378} {\bibfield
  {journal} {\bibinfo  {journal} {IEEE Transactions on Industry Applications}\
  }\textbf {\bibinfo {volume} {35}},\ \bibinfo {pages} {1162} (\bibinfo {year}
  {1999})}\BibitemShut {NoStop}%
\bibitem [{\citenamefont {Jackson~Kimball}\ \emph {et~al.}(2016)\citenamefont
  {Jackson~Kimball}, \citenamefont {Dudley}, \citenamefont {Li}, \citenamefont
  {Thulasi}, \citenamefont {Pustelny}, \citenamefont {Budker},\ and\
  \citenamefont {Zolotorev}}]{DMshielding}%
  \BibitemOpen
  \bibfield  {author} {\bibinfo {author} {\bibfnamefont {D.~F.}\ \bibnamefont
  {Jackson~Kimball}}, \bibinfo {author} {\bibfnamefont {J.}~\bibnamefont
  {Dudley}}, \bibinfo {author} {\bibfnamefont {Y.}~\bibnamefont {Li}}, \bibinfo
  {author} {\bibfnamefont {S.}~\bibnamefont {Thulasi}}, \bibinfo {author}
  {\bibfnamefont {S.}~\bibnamefont {Pustelny}}, \bibinfo {author}
  {\bibfnamefont {D.}~\bibnamefont {Budker}},\ and\ \bibinfo {author}
  {\bibfnamefont {M.}~\bibnamefont {Zolotorev}},\ }\href
  {https://doi.org/10.1103/PhysRevD.94.082005} {\bibfield  {journal} {\bibinfo
  {journal} {Phys. Rev. D}\ }\textbf {\bibinfo {volume} {94}},\ \bibinfo
  {pages} {082005} (\bibinfo {year} {2016})}\BibitemShut {NoStop}%
\bibitem [{\citenamefont {Stadnik}\ and\ \citenamefont
  {Flambaum}(2015{\natexlab{a}})}]{Stadnik2015}%
  \BibitemOpen
  \bibfield  {author} {\bibinfo {author} {\bibfnamefont {Y.~V.}\ \bibnamefont
  {Stadnik}}\ and\ \bibinfo {author} {\bibfnamefont {V.~V.}\ \bibnamefont
  {Flambaum}},\ }\href {https://doi.org/10.1103/PhysRevLett.114.161301}
  {\bibfield  {journal} {\bibinfo  {journal} {Phys. Rev. Lett.}\ }\textbf
  {\bibinfo {volume} {114}},\ \bibinfo {pages} {161301} (\bibinfo {year}
  {2015}{\natexlab{a}})}\BibitemShut {NoStop}%
\bibitem [{\citenamefont {Stadnik}\ and\ \citenamefont
  {Flambaum}(2015{\natexlab{b}})}]{Stadnik2015a}%
  \BibitemOpen
  \bibfield  {author} {\bibinfo {author} {\bibfnamefont {Y.~V.}\ \bibnamefont
  {Stadnik}}\ and\ \bibinfo {author} {\bibfnamefont {V.~V.}\ \bibnamefont
  {Flambaum}},\ }\href {https://doi.org/10.1103/PhysRevLett.115.201301}
  {\bibfield  {journal} {\bibinfo  {journal} {Phys. Rev. Lett.}\ }\textbf
  {\bibinfo {volume} {115}},\ \bibinfo {pages} {201301} (\bibinfo {year}
  {2015}{\natexlab{b}})}\BibitemShut {NoStop}%
\bibitem [{\citenamefont {Stadnik}\ and\ \citenamefont
  {Flambaum}(2016)}]{Stadnik2016}%
  \BibitemOpen
  \bibfield  {author} {\bibinfo {author} {\bibfnamefont {Y.~V.}\ \bibnamefont
  {Stadnik}}\ and\ \bibinfo {author} {\bibfnamefont {V.~V.}\ \bibnamefont
  {Flambaum}},\ }\href {https://doi.org/10.1103/PhysRevA.94.022111} {\bibfield
  {journal} {\bibinfo  {journal} {Phys. Rev. A}\ }\textbf {\bibinfo {volume}
  {94}},\ \bibinfo {pages} {022111} (\bibinfo {year} {2016})}\BibitemShut
  {NoStop}%
\bibitem [{\citenamefont {Hees}\ \emph {et~al.}(2018)\citenamefont {Hees},
  \citenamefont {Minazzoli}, \citenamefont {Savalle}, \citenamefont {Stadnik},\
  and\ \citenamefont {Wolf}}]{Hees2018}%
  \BibitemOpen
  \bibfield  {author} {\bibinfo {author} {\bibfnamefont {A.}~\bibnamefont
  {Hees}}, \bibinfo {author} {\bibfnamefont {O.}~\bibnamefont {Minazzoli}},
  \bibinfo {author} {\bibfnamefont {E.}~\bibnamefont {Savalle}}, \bibinfo
  {author} {\bibfnamefont {Y.~V.}\ \bibnamefont {Stadnik}},\ and\ \bibinfo
  {author} {\bibfnamefont {P.}~\bibnamefont {Wolf}},\ }\href
  {https://doi.org/10.1103/PhysRevD.98.064051} {\bibfield  {journal} {\bibinfo
  {journal} {Phys. Rev. D}\ }\textbf {\bibinfo {volume} {98}},\ \bibinfo
  {pages} {064051} (\bibinfo {year} {2018})}\BibitemShut {NoStop}%
\bibitem [{\citenamefont {Grote}\ and\ \citenamefont
  {Stadnik}(2019)}]{Stadnik2019}%
  \BibitemOpen
  \bibfield  {author} {\bibinfo {author} {\bibfnamefont {H.}~\bibnamefont
  {Grote}}\ and\ \bibinfo {author} {\bibfnamefont {Y.~V.}\ \bibnamefont
  {Stadnik}},\ }\href {https://doi.org/10.1103/PhysRevResearch.1.033187}
  {\bibfield  {journal} {\bibinfo  {journal} {Phys. Rev. Research}\ }\textbf
  {\bibinfo {volume} {1}},\ \bibinfo {pages} {033187} (\bibinfo {year}
  {2019})}\BibitemShut {NoStop}%
\bibitem [{\citenamefont {Kim}\ and\ \citenamefont {Perez}(2022)}]{Kim2022}%
  \BibitemOpen
  \bibfield  {author} {\bibinfo {author} {\bibfnamefont {H.}~\bibnamefont
  {Kim}}\ and\ \bibinfo {author} {\bibfnamefont {G.}~\bibnamefont {Perez}},\
  }\href@noop {} {\bibinfo {title} {{Oscillations of atomic energy levels
  induced by QCD axion dark matter}}} (\bibinfo {year} {2022}),\ \Eprint
  {https://arxiv.org/abs/2205.12988} {arXiv:2205.12988 [hep-ph]} \BibitemShut
  {NoStop}%
\bibitem [{\citenamefont {Banerjee}\ \emph {et~al.}(2022)\citenamefont
  {Banerjee}, \citenamefont {Perez}, \citenamefont {Safronova}, \citenamefont
  {Savoray},\ and\ \citenamefont {Shalit}}]{Banerjee2022}%
  \BibitemOpen
  \bibfield  {author} {\bibinfo {author} {\bibfnamefont {A.}~\bibnamefont
  {Banerjee}}, \bibinfo {author} {\bibfnamefont {G.}~\bibnamefont {Perez}},
  \bibinfo {author} {\bibfnamefont {M.}~\bibnamefont {Safronova}}, \bibinfo
  {author} {\bibfnamefont {I.}~\bibnamefont {Savoray}},\ and\ \bibinfo {author}
  {\bibfnamefont {A.}~\bibnamefont {Shalit}},\ }\href@noop {} {\bibinfo {title}
  {{The phenomenology of quadratically coupled ultra light dark matter}}}
  (\bibinfo {year} {2022}),\ \Eprint {https://arxiv.org/abs/2211.05174}
  {arXiv:2211.05174 [hep-ph]} \BibitemShut {NoStop}%
\end{thebibliography}
\end{document}